\documentclass[a4paper,10pt,final,oneside]{ar2e}

\usepackage{graphicx}% Include figure files
\usepackage{epsfig}% Include figure files
\usepackage{dcolumn}% Align table columns on decimal point
\usepackage{bm}% bold math
\usepackage{color}%
\graphicspath{{../Figures-EPS/}}
\begin{document}

\jname{Annual Review of Nuclear and Particle Science}
\jyear{2010}
\jvol{60}
%\ARinfo{Draft, \today}
%
\title{LUNA: Nuclear Astrophysics Deep Underground}
\markboth{Underground Nuclear Astrophysics}{Underground Nuclear Astrophysics}
\author{%
Carlo Broggini\thanks{Corresponding author}
	\affiliation{Istituto Nazionale di Fisica Nucleare (INFN), Sezione di Padova, Padova, Italy\\
	broggini@pd.infn.it}
Daniel Bemmerer
	\affiliation{Forschungszentrum Dresden-Rossendorf, Dresden, Germany\\
	d.bemmerer@fzd.de}
Alessandra Guglielmetti
	\affiliation{Universit\`a degli studi di Milano and \\ Istituto Nazionale di Fisica Nucleare (INFN), Sezione di Milano, Milano, Italy\\
	alessandra.guglielmetti@mi.infn.it}
Roberto Menegazzo
	\affiliation{Istituto Nazionale di Fisica Nucleare (INFN), Sezione di Padova, Padova, Italy\\
	roberto.menegazzo@pd.infn.it}
}
\date{\today}

\begin{abstract}
Nuclear astrophysics strives for a comprehensive picture of the nuclear reactions responsible for synthesizing the chemical elements and for powering the stellar evolution engine. Deep underground in the Gran Sasso laboratory the cross sections of the key reactions of the proton-proton chain and of the Carbon-Nitrogen-Oxygen (CNO) cycle have been measured right down to the energies of astrophysical interest. The salient features of underground nuclear astrophysics are summarized here. The main results obtained by LUNA in the last twenty years are reviewed, and their influence on the comprehension of the properties of the neutrino, of the Sun and of the Universe itself are discussed. Future directions of underground nuclear astrophysics towards the study of helium and carbon burning and of stellar neutron sources in stars are pointed out.
\end{abstract}

\maketitle

% =======================
\section{Introduction}
The stars which like diamonds fill the sky at night fascinating our mind are not perfect and everlasting bodies as believed by the ancient philosophers.
On the contrary, gravity triggers the birth of a star which then works as a more or less turbulent chemical factory \cite{Eddington20-Nature} to finally die out in a quiet or violent way, depending on its initial mass \cite{Clayton84-Book}. As a matter of fact, only hydrogen, helium and lithium are synthesized in the first minutes after the big-bang. All the other elements 
of the periodic table are produced in the thermonuclear reactions taking place inside the stars, where the nuclear roots of life itself are embedded in \cite{Rolfs88-Book,Clayton03-Book}. The aim of nuclear astrophysics is to reach a comprehensive picture of all these reactions  which realize the transmutation of the chemical elements and which provide the energy to run the engine of stellar evolution \cite{Iliadis07-Book}.

The knowledge of the reaction cross-section at the stellar energies lies at the heart of nuclear astrophysics.
At these energies the cross sections are extremely small. Such smallness makes the star life-time of the length we observe, but it also makes impossible the direct measurement in the laboratory. The rate of the reactions, characterized by a typical energy release of a few MeV, is too low, down to a few events per year, in order to stand out from the background. LUNA, Laboratory for Underground Nuclear Astrophysics, started twenty years ago to run nuclear physics experiments in an extremely low-background environment: the Gran Sasso Laboratory. Since you cannot distinguish the timbre of a piano note in a crowded circus, but you can do so if you are in a music hall, then the LUNA physicists are tuning their accelerators and detectors in the 'music hall' deeply inside the mountain 'to 
listen' to the tiny signal from nucleo-synthesis reactions, reproducing this way  in the laboratory what Nature makes inside the stars.

In this review the main features of thermonuclear reactions at very low energy, the characteristics of the background attainable in Gran Sasso and the experimental apparata employed by LUNA
will first be described. Then, an overview of hydrogen burning in stars will be given and the LUNA main results will be discussed, with  emphasis on their impact on the picture of the Sun and of the neutrino. Finally, the next steps of underground nuclear astrophysics, mainly devoted to the study of the helium and carbon burning and of the neutron sources in stars, will be outlined.
% =======================
\section{Thermonuclear reactions}
Thermouclear reactions between nuclei occur inside the star in a relatively narrow energy window, placed at an energy much lower than the height of the barrier arising from the Coulomb repulsion between nuclei. As a consequence, the reaction can only take place owing to the quantum mechanical tunnel effect
which leads to a very small, but not vanishing, probability for the incoming nucleus to penetrate the Coulomb barrier and to reach its reaction partner.

Because of the tunnel effect, at these energies the reaction cross-section 
$\sigma(\mathrm{E})$
drops almost exponentially 
with decreasing energy $E$:
\begin{equation}
\sigma(E)= \frac{S(E)}{E}\,{exp(-2\,\pi\, \eta)} \label{yielddef1}
\end{equation}
where $S(E)$ is the so-called astrophysical $S$-factor and
$2\, \pi\, \eta=31.29\, Z_1\, Z_2(\mu/E)^{1/2}$.
$Z_1$ and $Z_2$  are  the electric
charges of the nuclei,
$\mu$ is the reduced mass (in a.m.u.),  and  $E$  is  the energy (in keV) in the center of mass system
\cite{Clayton84-Book,Rolfs88-Book}. For most of the reactions, the astrophysical $S$-factor varies only slowly with energy and contains all the nuclear physics information.

The reaction rate in the hot plasma of a star, with temperatures in the range of tens to hundreds of millions Kelvin, 
is obtained by weighting the reaction cross section $\sigma (E)$ with the energy distribution of the colliding nuclei: a Maxwell-Boltzmann $\phi(E)$ peaked at energies of 1-10\,keV.
The product between  $\sigma(E)$ and $\phi(E)$ identifies the energy window where the reactions occur in the star: the Gamow peak. At lower energies the cross section is too small whereas at higher energies the nuclei in the tail of the Maxwell-Boltzmann are too few.
Finally, the energy balance of the nuclear reaction is determined by the Q-value, which corresponds to the mass difference between the entrance and exit channels.

In order to obtain the precise nuclear physics data required by modern astrophysics one should measure the relevant reaction cross section directly at the energy of the astrophysical scenario to be studied.
For the solar reactions
the cross-section values at the Gamow peak range from picobarn to femtobarn and even below (1 barn = 10$^{-24}$\,cm$^2$). These extremely low cross sections, while trying the patience of the physicists with ultra-low count rates, are a blessing to mankind in general, because they allow hydrogen burning to proceed in the Sun at a placid pace for several billions of years to come. 

However, the low count rates pose another experimental problem: in direct laboratory measurements at the Earth's surface the signal to background ratio is too small. So, instead of a direct measurement, the observed energy dependence of the cross-section at high energies is extrapolated to the low energy region. However, any extrapolation is fraught with uncertainty.
For example, there might be narrow resonances at low energy or even resonances below the reaction threshold that influence the cross section at the Gamow peak. Those effects cannot always be accounted for by an extrapolation.

In addition, another effect can be studied at low energies: electron
screening \cite{Salpeter54-AustralJP,Assenbaum87-ZPA,Raiola04-EPJA,Huke08-PRC}. In the laboratory, the electron cloud surrounding the target nucleus 
partially screens its positive electric charge. This reduces the height of the 
Coulomb barrier, thus increasing the tunneling probability and eventually the reaction cross-section. The screening effect has to be taken into 
account in order to derive the cross-section for bare nuclei, which is the input data to nucleosynthesis network codes. These codes, in turn, have to take into account the screening by electrons in the stellar plasma \cite{Salpeter54-AustralJP}.

All these effects mean that, despite the impressive-sounding temperatures of many millions Kelvin in stellar interiors, actually the study of the star requires nuclear physics experiments at very low energy, measuring exceedingly small cross sections.

% =======================
\section{Underground nuclear astrophysics}

For stable stellar hydrogen burning, the relevant temperatures range from 20 to 100 $\cdot$ 10$^6$\,K, corresponding to Gamow peak energies of 10--50\,keV, depending on the precise reaction of interest.
The challenge of the measurement at the Gamow peak is to suppress the laboratory and beam induced background and then to enhance the signal by boosting the beam intensity, target density, and detection efficiency.

\subsection{Laboratory background}

\begin{figure}[tb]
\centering
 \includegraphics[angle=0,height=0.8\textheight,angle=-90,width=1.0\textwidth]{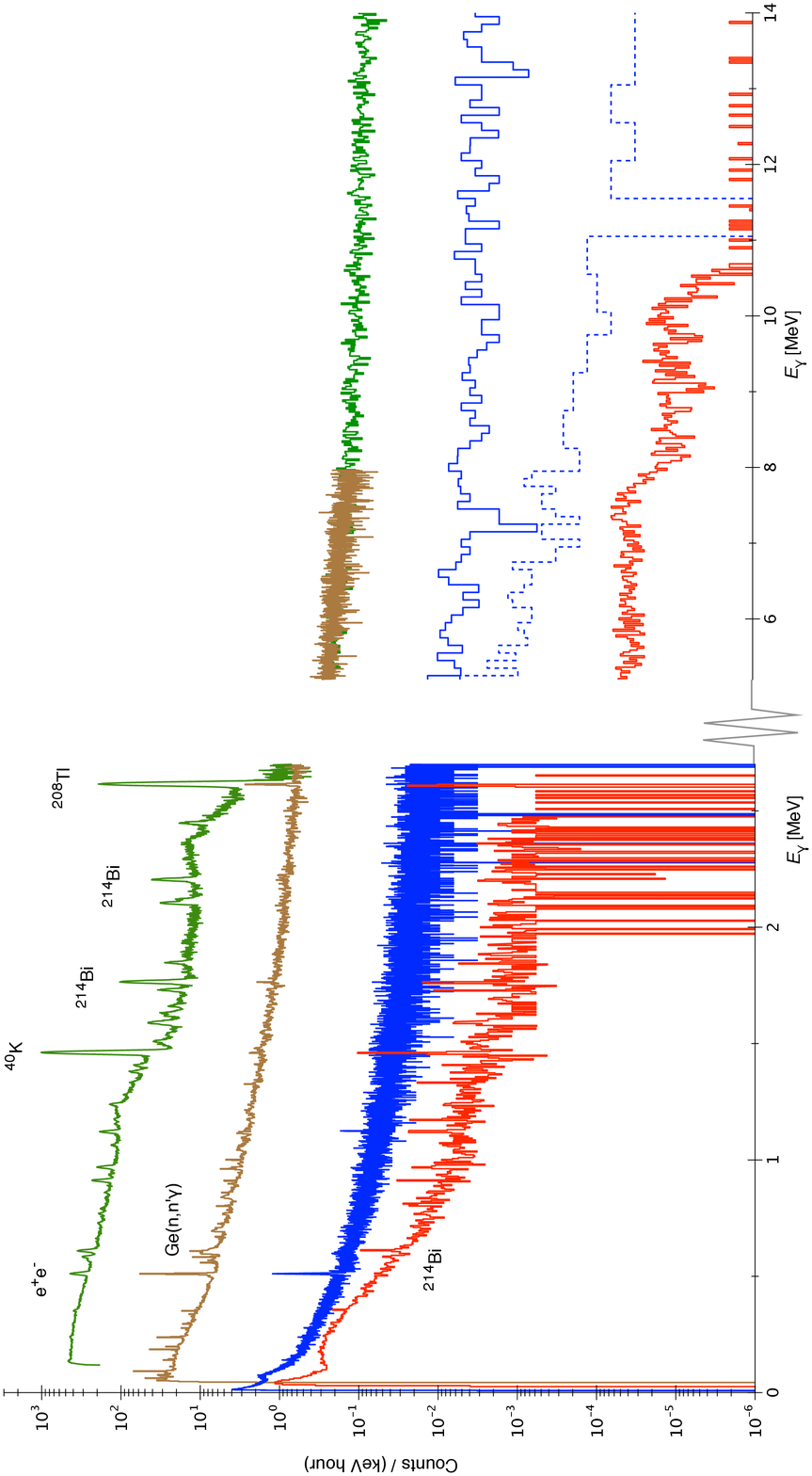}
 \caption{\label{fig:figure1} Laboratory $\gamma$-ray background measured with a 100\% relative efficiency germanium detector 
at $E_\gamma$ $<$ 2.7\,MeV and with a BGO detector (scaled for equal volume with the germanium) for 5.2\,MeV $<$ $E_\gamma$ $<$ 14\,MeV. 
Green: Earth's surface, no shield. Brown: Earth's surface, lead shield. Blue: 110 m.w.e. underground Felsenkeller laboratory (Dresden), lead shield \cite{Koehler09-Apradiso}. Blue dashed line: actively vetoed spectrum in the 110 m.w.e. underground Felsenkeller lab. Red: 3800 m.w.e. LUNA lab, lead shield for the germanium \cite{Caciolli09-EPJA}, no lead shield for the BGO \cite{Bemmerer05-EPJA}.}
\end{figure}

The laboratory $\gamma$-background has two main sources: natural and cosmic-ray induced radioactivity.
The decay of radioisotopes from the natural decay chains are evident with many characteristic $\gamma$-lines in the region below 2.6\,MeV, where the highest energy $\gamma$ due to natural radioactivity is found. Figure~\ref{fig:figure1} illustrates the different steps in background reduction that are necessary for nuclear astrophysics experiments. 

From a naked detector to a commercially available graded shield \cite{Gilmore08-Book}, including 10\,cm lead, there is already a sizable reduction. This is due to the suppression of the soft component of cosmic-rays  and of the $\gamma$-rays from nearby radioisotopes. By going to a shallow underground laboratory at ~100\,m.w.e. (meters water equivalent) and applying a more sophisticated shield, another reduction is possible \cite{Koehler09-Apradiso}. At this point the remaining background is dominated by muon induced events. 

These muon induced events stem from 
the decay of the radioactive nuclei due to muon spallation and to the capture of stopping negative muons, from the inelastic scattering and capture of the muon induced neutrons, from   
the energy loss of muons passing through the detector. The prompt effects of the muons can be reduced by surrounding the detector with a veto counter. On the contrary, the muon induced radioactivity may have a relatively long life-time (1-10 s), so a veto counter will not help against it. 
Instead, this remaining background can be overcome by moving to a deep underground laboratory \cite{Caciolli09-EPJA,Laubenstein04-Apradiso}, where the muon flux is strongly reduced (e.g. by six orders of magnitude \cite{MACRO90-PLB} at the 3800 m.w.e. deep Gran Sasso laboratory).

At higher $\gamma$-ray energies, i.e. $E_\gamma$ $>$ 2.6\,MeV, these considerations become more transparent. In this $\gamma$-energy region
no significant improvement in background can be reached by applying an additional lead shield. Instead, going to a shallow-underground laboratory helps somewhat, and a muon veto reduces the background. When going deep underground, the muon flux becomes negligible, and the veto does not further reduce the background counting rate. A further reduction below this very low $\gamma$-background observed at LUNA \cite{Bemmerer05-EPJA,Szucs10-EPJA} can in principle be achieved by shielding the set-up against neutrons. The neutron flux in Gran Sasso is already reduced by three orders of magnitude as compared to the outside and it is due to the ($\alpha$,n) reactions in the rock \cite{Wulandari04-APP}.

For the sake of completeness, we remind that
nuclear astrophysics is just one of the fields explored deep underground.  The salient features of underground physics have been reviewed in the past \cite{Formaggio04-ARNPS}. Major underground activities are 
the detection of neutrinos generated by the hydrogen burning in the Sun \cite{Bethe39-PR_letter,Weizsaecker38-PZ}, by the radioactivity in the Earth \cite{KamLAND05-Nature} and by cosmic-rays in the Earth's atmosphere \cite{Gaisser02-ARNPS},
the search for rare processes such as neutrinoless double-$\beta$ decay \cite{Elliott02-ARNPS}, proton decay \cite{Perkins84-ARNPS} and dark matter interactions\cite{Gaitskell04-ARNPS}.  Recently a compendium of major facilities has been presented \cite{Bettini07-arxiv}.

\subsection{Beam induced background}

In any experiment with an ion beam, the beam interacts also with nuclei other than the target to be studied. Such nuclei could be found in the beam transport system (mainly beam limiting apertures, but also drift tubes, magnets and residual gas) but also as parts of the target itself. Such interactions can give rise to $\gamma$-ray background. 

This ion beam induced background is clearly independent of the underground depth and must be dealt with by reducing the inventory of materials hit by the ion beam and by appropriate precautions to eliminate particularly worrisome components. For low-energy nuclear astrophysics experiments this task is greatly facilitated by two aspects. Firstly, contaminants with atomic number greater than the target to be studied will generally have lower interaction probability than the target itself because of their higher Coulomb barrier. Many experiments study light nuclei of $Z$ $\leq$ 8, so common materials such as aluminum or steel will generally not give background. Secondly, the otherwise commonly found activation of beamline components is not an issue with low-energy proton or $\alpha$-beam. In cases where the background still plays a role, it must be carefully identified, localized \cite[e.g.]{Bemmerer05-EPJA} and eliminated. 

\subsection{LUNA at Gran Sasso}

There are several possible techniques to measure cross sections at the Gamow peak. All of them have so far been pursued just in one laboratory: LUNA. 

For nuclear reactions where charged particles are emitted, generally an in-beam measurement at the surface of the Earth is possible. This is true except for cases where coincident background due to cosmic rays is a problem. If so, only an underground measurement is feasible, as it has been demonstrated for the $^3$He($^3$He,2p)$^4$He reaction \cite{Junker98-PRC,Bonetti99-PRL}. 

For nuclear reactions where only $\gamma$-rays are emitted, there are in principle two approaches. In-beam $\gamma$-spectrometry deep underground has been applied to the studies of the $^2$H(p,$\gamma$)$^3$He \cite{Casella02-NPA}, $^{14}$N(p,$\gamma$)$^{15}$O \cite{Formicola04-PLB,Imbriani05-EPJA,Lemut06-PLB,Bemmerer06-NPA,Marta08-PRC}, $^3$He($^4$He,$\gamma$)$^7$Be \cite{Confortola07-PRC,Costantini08-NPA} and $^{15}$N(p,$\gamma$)$^{16}$O \cite{Bemmerer09-JPG} reactions. 

The in-beam approach has to take some uncertainty into account due to the usually not well known angular distribution of the emitted $\gamma$-rays. This weakness can be overcome in selected cases where the created nucleus is radioactive, thus allowing an independent cross-check. An activation study with deep underground activity-counting has been performed in the study of $^3$He($^4$He,$\gamma$)$^7$Be \cite{Bemmerer06-PRL,Gyurky07-PRC}. All these approaches benefit from the reduced background deep underground.
% =======================
\section{Experimental apparata at LUNA}
The measurement of the cross section and the determination of the
astrophysical S-factor for thermonuclear reactions require an experimental apparatus basically
composed of an accelerator, a target and a detection system. 
\subsection{Accelerators}
\label{sec:expapparata_accelerators}
Two different accelerators have been used at LUNA: a compact 50 kV "home-made" machine
\cite{Greife94-NIMA} and a commercial 400 kV one \cite{Formicola03-NIMA}. Common features
of the two accelerators are the high beam current, the long term stability and the precise
beam energy determination. The first feature is required to maximize the reaction
rate, the second is due to the long time typically needed for a cross section measurement,
while the third is important because of 
the exponential-like energy dependence of the cross section. 

The 50 kV machine was designed and built at the Ruhr Universit\"at in Bochum and then
moved to Gran Sasso. It consisted of a duoplasmatron ion source, an ion beam
extraction and acceleration system and a double-focusing 90$^\circ$ analysing magnet. Its compact shape optimized the beam transmission. All in all the machine
\cite{Greife94-NIMA} was able to deliver beams of protons, $^3$He$^{+}$ and $^4$He$^{+}$ of 300-500
$\mu$A at energies between 10 and 50 keV with an energy spread of  less than 20 eV and
a long term stability. This allowed the study of two fundamental reactions of
the p-p chain: $^3$He($^3$He,2p)$^4$He and $^2$H(p,$\gamma$)$^3$He at solar Gamow peak
energies and of the screening effect  in the $^2$H($^3$He,p)$^4$He
reaction. 

\begin{figure}
\centerline{\includegraphics[height=0.45\textheight]{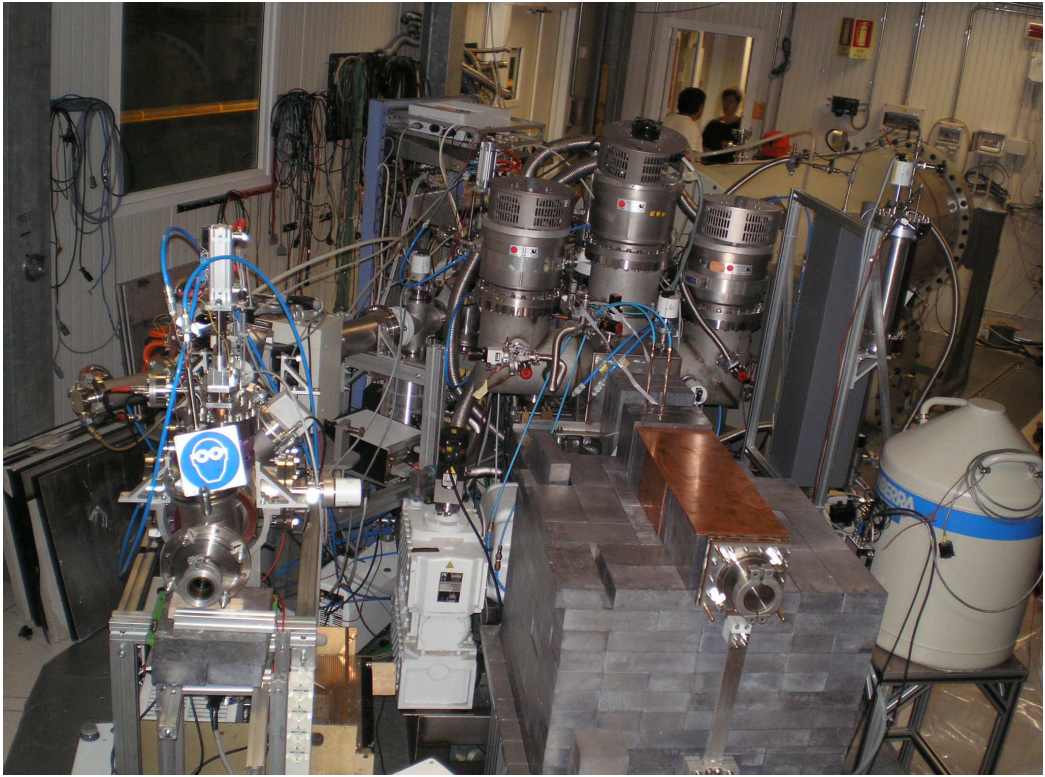}}
\caption{The LUNA set-up with the two different beam lines in the foreground and the accelerator in the back. The beam line to the left is dedicated to the measurements with solid target whereas the one on the right hosts the windowless gas target. The set-up for the study of $^3$He($^4$He,$\gamma$)$^7$Be is shown during installation with the shield only partially mounted.}
\label{fig:figure2}
\end{figure}

Even though it produced outstanding results \cite{Bonetti99-PRL,
Casella02-NPA, Costantini00-PLB}, this machine should be considered as a pilot project
towards a higher energy facility, namely the 400 kV accelerator
\cite{Formicola03-NIMA} shown in Fig. \ref{fig:figure2}. This electrostatic
accelerator is embedded in a tank filled with a mixture of N$_2$/CO$_2$ gas at 20 bar, working as insulator.
The high voltage is generated by an Inline-Cockroft-Walton power supply located inside
the tank. The radio-frequency ion source, directly mounted on the accelerator tube can
provide beams of 1 mA hydrogen and 500\,$\mu$A He$^{+}$ over a continuous operating time of 40
days. The ions can be sent into one of two different and parallel beam lines, this way allowing the installations of two different target set-ups. In the
energy range 150-400 keV, the accelerator can provide up to 500 $\mu$A of protons and
250\,$\mu$A of alphas at the target stations, with 0.3\,keV accuracy on the beam energy,
100\,eV energy spread and 5\,eV/h long-term stability. Finally, the accelerator is controlled by
PLC-based computers which allow for a safe operation over long periods without an operator present in situ.

\subsection{Targets and ancillary measurements}
\begin{figure}
\centerline{\includegraphics[width=\textwidth,height=0.65\textheight,keepaspectratio,angle=-90]{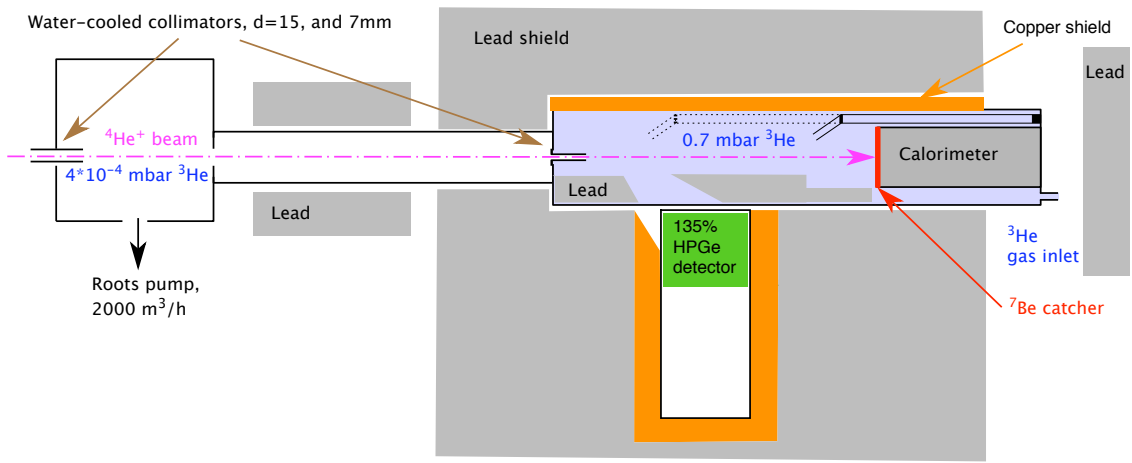}}
\caption{The set-up for the study of $^3$He($^4$He,$\gamma$)$^7$Be. The device to detect 
Rutherford scattering, the calorimeter and the germanium detector are indicated.}
\label{fig:figure3}
\end{figure}

LUNA measurements have been performed either with solid or gas targets. Solid
targets may contain a larger number of atoms per cm$^2$ with respect to gas targets: typical areal densities are in the range 2$\cdot$10$^{17}$-2$\cdot$10$^{18}$. They also 
allow for the measurement of the cross-section angular dependence since the beam-hit
position on the target can be rather precisely determined. Gas targets, instead, are more
stable to beam bombardment and may reach an extreme purity, essential to minimize the beam
induced background. In this case areal densities used in LUNA are between 5$\cdot$10$^{16}$ and 1$\cdot$10$^{18}$. While for solid targets LUNA has
taken advantage of well known production techniques such as implantation, evaporation
or sputtering \cite{Imbriani05-EPJA}, for gas targets a specific set-up,
consisting of a windowless gas target (i.e a differentially pumped target) with
recirculation system \cite{Casella02-NIMA, Gyurky07-PRC},
has been
designed and installed at Gran Sasso 
and is partially shown in Fig.\ref{fig:figure3}. 

A typical disadvantage of a gas
target is the so called "beam heating" effect, due to the power deposition which gives rise to the heating and the thinning of the gas 
along the beam path. Different techniques 
are necessary to obtain the real
number of atoms per cm$^2$ in the beam region. A specific
set-up based on the Rutherford scattering was designed \cite{Marta06-NIMA} and it allowed for a systematic
uncertainty on the gas density determination due to the beam heating effect below 2\%
\cite{Bemmerer06-PRL}.
Alternatively, the target thickness along the beam can be obtained by the energy shift of a proper well
known resonance \cite{Goerres80-NIM,Bemmerer06-NPA}. 

Solid targets, in turn, may rapidly
deteriorate with the intense beam impinging onto, thus changing the density profile
which is an essential ingredient of the reaction rate. In order to monitor the target
thickness and its stability under beam, the resonance scan technique may be used. This
consists in selecting a resonance of the reaction to be studied or of a parasitic
reaction and measuring the yield
profile as a function of the beam energy. The selected resonance should be
characterized by an energy width smaller or equal to the target thickness. The target thickness as well as the maximum yield can be repeatedly monitored
during the whole measurement time. 

For the cross section measurement it is essential to know the beam current on target.
While for a solid target the current is measured through a conventional Faraday cup, for a gas
target this is not possible due to the ion charge state changes
\cite{Allison58-RMP}. Therefore, LUNA has developed a beam calorimeter
\cite{Casella02-NIMA} with a constant temperature gradient between a hot and a cold
side. The power to the hot side is provided by the beam and by resistors
embedded in the end-cap of the calorimeter.
The number of projectiles is then obtained by the difference in heating powers dissipated by the resistors
with and without beam divided by the kinetic energy of the projectiles themselves at
the calorimeter surface. If properly calibrated, the calorimeter can give a systematic
uncertainty of the order of 1\% on the beam current \cite{Casella02-NIMA}.
Calorimeter calibration is an example 
of the measurements done in laboratories at the Earth surface and important to complete and integrate the results obtained underground.
\subsection{Detectors}
\label{sec:expapparata_detectors}
Apart from the $^3$He($^3$He,2p)$^4$He reaction where protons had to
be detected and this was accomplished by commercial silicon detectors as single stage
or $\Delta$E-E telescopes, all the other LUNA measurements required the detection of
$\gamma$-rays. Here the choice of the most suitable detector depends on the desired
physical information. The 4$\pi$ BGO summing crystal used at LUNA \cite{Casella02-NIMA},
a cylinder 28 cm long with a 20 cm diameter and a 6 cm bore-hole,
can reach an efficiency as high as 70\% for 7 MeV $\gamma$-ray thus allowing the measurement of
extremely low reaction yields. As a
counterpart, the BGO energy resolution is very poor and does not allow measurements of
cascades and branching ratios to different levels since most of the $\gamma$ ray
transitions are summed in a single peak. With a germanium detector the efficiency
dramatically decreases to the level of a few permille but the energy resolution is much better thus
allowing to disentangle complex gamma cascades. Moreover, angular
distribution measurements are possible by moving the detector to different angles with respect to the ion beam.
Nevertheless, summing effects can disturb the data
evaluation for close geometry configurations. This can be addressed by using a composite detector like a Clover \cite{Elekes03-NIMA}, i.e. 4 small germanium detectors closed inside the same cryostat, this way reducing 
summing-in corrections. 

% =======================
\section{Hydrogen burning in the Sun}

As the solar mass contracted from an initially large gas cloud, half of the gravitational energy 
released has been radiated into the space and half converted into kinetic energy of the hydrogen
and helium nuclei, thus increasing the temperature of the system \cite{Clayton84-Book}. This way the solar mass was loosing energy,
contracting and heating up. If gravitational energy were the only energy source then the Sun
would have ended its life when only 30 million years old, as estimated by Lord Kelvin in the 19th century.
On the contrary, at the central temperature of about 10 million degrees the kinetic energy of the hydrogen nuclei was high enough
to penetrate with significant probability the Coulomb barrier and to switch on the hydrogen burning: 4$^{1}$H$\Rightarrow$$^{4}$He+2e$^{+}$+2$\nu$$_{e}$,
i.e. the fusion of hydrogen into helium with the production of positrons and neutrinos. The mass of the helium nucleus is lower
then 4 times the proton mass, as a consequence about 0.7$\%$ of the hydrogen rest mass is converted into energy in each 
of the transmutations.

Hydrogen fusion supplies the energy necessary to halt the contraction and it provides all the energy
required for the long and quiet life of the star. The Sun is a middle-aged main sequence star which shines by burning hydrogen fuel.
It started burning 
hydrogen about 4.5 billion years ago and, in about 5 billion years, it will burn helium to finally end-up as a 
celestial body mainly consisting of carbon and oxygen. 
In the central region of the Sun, at the temperature of 15 million degrees and with a density of about 150 gr cm$^{-3}$,
the hydrogen burning does not take place in one step only but it proceeds through series of two body reactions:
the proton-proton (pp) chain and the CNO cycle. The relative importance of the different reactions is
determined by the abundance of the nuclear species which are fusing together and by their fusion cross section at the Gamow peak energy.

\subsection{proton-proton chain}

\begin{figure}%
\centerline{
\includegraphics[width=0.75\linewidth]{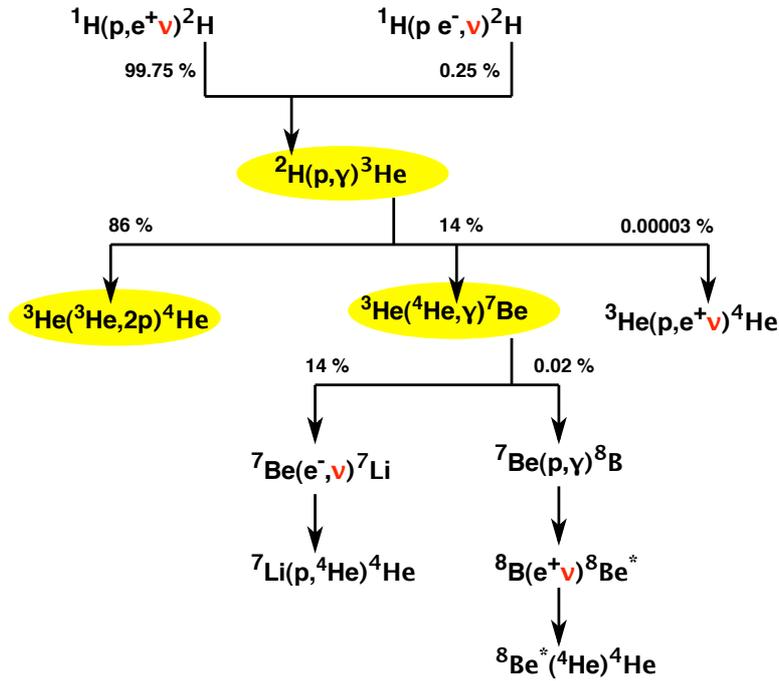}
}
\caption{The proton-proton chain. The reactions studied by LUNA are highlighted in yellow.}
\label{fig:figure4}
\end{figure}

The first step and bottleneck of the chain is the production of deuterium \cite{Bahcall89-Book}. It takes place through two different weak processes
(Fig. \ref{fig:figure4}):
$^{1}$H(p,$e^{+}\nu$)$^{2}$H (giving rise to the so called pp neutrinos) and $^{1}$H(p$e^{-}$,$\nu$)$^{2}$H (pep neutrinos). 
The latter, being a three body process, is strongly suppressed (about a factor 400) as compared to the former. 
Once produced, the deuterium quickly burns with hydrogen to synthesize $^{3}$He. At this point a complex and rich scenario opens 
with several possible branches. $^{3}$He(p,$e^{+}\nu$)$^{4}$He (hep neutrinos) has a negligible rate since it is a weak process 
further suppressed, as compared to $^{1}$H(p,$e^{+}\nu$)$^{2}$H, because of the atomic number of helium. The most probable fate
of $^{3}$He is to fuse with another $^{3}$He nucleus to finally produce $^{4}$He in the strong reaction $^{3}$He($^{3}$He,2p)$^{4}$He.
In about 14$\%$ of the cases the fusion takes place with the much more abundant $^{4}$He through the electromagnetic process 
$^{3}$He($^{4}$He,$\gamma$)$^{7}$Be. At this level the chain branches again due to the competition between the electron 
capture decay of $^{7}$Be: $^{7}$Be($e^{-}$,$\nu$)$^{7}$Li ($^{7}$Be neutrinos) and the fusion of $^{7}$Be with hydrogen:
$^{7}$Be(p,$\gamma$)$^{8}$B. The former is a weak process which is about a factor thousand more probable than the latter since it has no Coulomb barrier suppression.
Once produced, $^{7}$Li quickly fuses with hydrogen to produce $^{8}$Be which is extremely unstable and splits into two helium
nuclei. $^{7}$Be(p,$\gamma$)$^{8}$B seldom occurs but it is of crucial 
importance since it leads to the 'high' energy neutrinos emitted in the $^{8}$B decay to $^{8}$Be.

\subsection{Carbon-Nitrogen-Oxygen cycle}

In the Carbon-Nitrogen-Oxygen cycle (CNO cycle) the conversion of hydrogen into helium is achieved with the aid of the carbon previously 
synthesized in older stars \cite{Rolfs88-Book}. Carbon works as a catalyst, it is not destroyed by the cycle and
 it strongly affects the rate of the CNO cycle with its abundance. 
Since $^{15}$N(p,$\gamma$)$^{16}$O has a cross section which is about a factor two thousand lower than $^{15}$N(p,$\alpha$)$^{12}$C,
the second CNO cycle is strongly suppressed as compared to the first one. In the Sun the CNO cycle accounts for just a small fraction of the nuclear
energy production (less than 1$\%$) and it is ruled by 
$^{14}$N(p,$\gamma$)$^{15}$O,
the bottleneck reaction.
Only at the temperature of 20 millions degree it would give the same contribution
as the pp chain, whereas it would dominate at higher temperatures, where the effects of the Coulomb barriers do not strongly affect anymore
the energy production rate. 

\subsection{Neutrinos}

Neutrinos are particle which interact weakly with matter, they travel at a speed which is essentially the speed of light
and they reach the Earth 8 minutes after their birth in the central region of the Sun. The calculation of their flux is straightforward 
when we know that hydrogen fusion
4$^{1}$H$\Rightarrow$$^{4}$He+2e$^{+}$+2$\nu$$_{e}$ is producing 26.7 MeV energy and if we make the much reasonable assumption that the present 
luminosity of the Sun corresponds to the present nuclear energy production rate \cite{Castellani97-PhysRep} (it takes more than 10$^{4}$ years
to the electromagnetic energy to reach the surface of the Sun). We only have to divide the solar luminosity, 3.85$\cdot$10$^{26}$ Watt, by the energy required to have one neutrino, 13.35 MeV
$\equiv$2.14$\cdot$10$^{-12}$ Joule, to obtain a rate of 1.80$\cdot$10$^{38}$ neutrinos per second. This corresponds to a flux of about 60 billions neutrinos per squared 
centimeter per second on the Earth. Instead, nuclear physics, in particular the cross section of the different reactions of
the pp chain and of the CNO cycle, is the key ingredient to calculate the energy spectrum of the solar neutrinos: the different branches give rise to neutrinos of different energy.

In particular, pp neutrinos have a flux of 6.04$\cdot$10$^{10}$ cm$^{-2}$s$^{-1}$ \cite{PenaGaray08-arxiv}, corresponding to 92$\%$ of the total neutrino flux.
Their continuous spectrum has the end-point energy of 0.42 MeV, which makes their detection extremely difficult. $^{7}$Be neutrinos are produced with two different 
energies: 0.86 MeV (89.7$\%$ branching ratio) and 0.38 MeV. The 0.86 MeV $^{7}$Be neutrinos are the second biggest component of the spectrum,
amounting to 7$\%$ of the total with a flux of 4.55$\cdot$10$^{9}$ cm$^{-2}$s$^{-1}$. $^{8}$B neutrinos have a much lower flux, 4.72$\cdot$10$^{6}$ cm$^{-2}$s$^{-1}$: fewer than one neutrino
over ten thousand is coming from the $^{8}$B decay. However, their relatively high end-point energy, about 15 MeV, makes their detection the least difficult one.
As a matter of fact, $^{8}$B neutrinos are the best studied neutrinos from the Sun so far. The CNO cycle is producing neutrinos with end-point energy of 1.20 MeV ($^{13}$N) and 
1.73 MeV ($^{15}$O). The latest results of the standard solar model \cite{PenaGaray08-arxiv} predicts a 0.5$\%$ contribution of the CNO neutrinos to the total flux.

In 1964 J.N. Bahcall and R. Davis Jr. proposed to detect solar neutrinos in order to see into the interior of the Sun
and thus directly verify the hypothesis of nuclear energy generation in stars \cite{Bahcall64-PRL,Davis64-PRL}. About 40 years of much refined experimental and theoretical
work have been required to show that the source of the energy radiated by the Sun is the hydrogen fusion in the solar interior. In addition, solar neutrinos told us something 
extremely important about the nature of neutrino itself: it oscillates. Produced as electron neutrino inside the Sun, it may be a muon or tau neutrino when
reaching the Earth.

% =======================
\section{Hydrogen burning studied at LUNA}

 LUNA started as a pilot project in the year 1991. In the following paragraphs we will discuss the brush-strokes given by LUNA 
 during the last 20 years to the current picture of the Sun and of the neutrino.

\subsection{$^2$H(p,$\gamma$)$^{3}$He: the energy source of the proto-star}

Inside the Sun, the $^2$H(p,$\gamma$)$^{3}$He reaction controls the 
equilibrium abundance of deuterium.
In a different scenario, $^2$H(p,$\gamma$)$^{3}$He 
is the reaction which rules the life of proto-stars before
they enter the main sequence phase. Proto-star models predict 
that a star forms by accretion of interstellar material onto a
small contracting core. Until the temperature remains below 10$^{6}$\,K, the main source of energy is the gravitational contraction.
When the temperature approaches 10$^{6}$\,K 
the first "nuclear fire"
is switched on inside the star: 
the primordial deuterium is 
converted into $^{3}$He
via $^2$H(p,$\gamma$)$^{3}$He, thus providing 5.5\,MeV for each reaction.
The total amount of nuclear energy generated by this d-burning is comparable
with the whole gravitational binding energy of the star. The on-set
of d-burning slows down the contraction, increases the lifetime
of the star and freezes its observational properties until
the original deuterium is fully consumed. A reliable 
knowledge of the rate of $^2$H(p,$\gamma$)$^{3}$He down to a few keV
(the Gamow peak in a proto-star) is a fundamental prerequisite for the proto-stellar models.

$^2$H(p,$\gamma$)$^{3}$He is also a cornerstone 
in the big-bang nucleosynthesis (BBN). Because of the deuterium 
"bottleneck" \cite{Weinberg72-Book}, i.e. the photo-disintegration of deuterium,
the formation of $^{3}$He is delayed until the temperature 
drops to about 8$\cdot$10$^{8}$\,K. Once again, the knowledge of 
the cross section at low energies is required.

\begin{figure}
\centerline{
\includegraphics[width=0.75\linewidth]{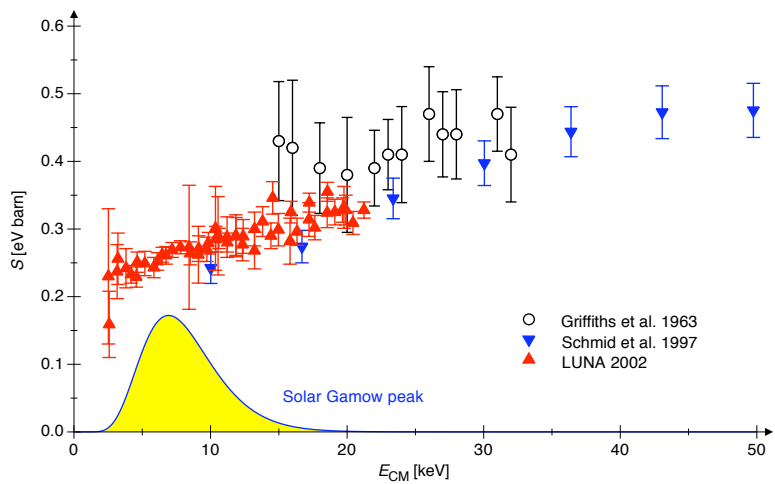}
}
\caption{The  $^2$H(p,$\gamma$)$^{3}$He astrophysical factor $S(E)$ with the total error.}
\label{fig:figure5}
\end{figure}

The $^2$H(p,$\gamma$)$^{3}$He cross section measurement was performed at LUNA
with the 50\,kV accelerator connected to a differentially
pumped gas-target system designed to fit the large
BGO $\gamma$-ray detector \cite{Casella02-NIMA}. The BGO, 
placed around the deuterium target, was detecting the 5.5\,MeV $\gamma$-ray with
70\% efficiency.
The LUNA results \cite{Casella02-NPA} are given in Fig. \ref{fig:figure5} 
together with 
two previous measurements 
\cite{Griffiths63-CJP,Schmid96-PRL} 
of the 
astrophysical factor $S(E)$ at low energy.
The agreement with the theoretical calculations is excellent
\cite{Marcucci06-NPA}.

\subsection{$^{3}$He($^{3}$He,2p)$^{4}$He: in search of the resonance}

The initial activity of LUNA has been focused 
on the $^{3}$He($^{3}$He,2p)$^{4}$He 
cross section measurement within the solar 
Gamow peak (15-27 keV).
Such a reaction is a key one of the pp chain.
A resonance at the thermal energy of the Sun 
was suggested 
long time ago  
\cite{Fowler72-NATURE,Fetisov72-PLB} to explain the observed $^{8}$B 
solar neutrino flux. Such a resonance 
would decrease the relative contribution of the alternative 
reaction  
$^{3}$He($^4$He,$\gamma$)$^{7}$Be, which generates the branch responsible for
$^{7}$Be and $^{8}$B
neutrino production in the Sun.

\begin{figure}
\centerline{
\includegraphics[width=0.75\linewidth]{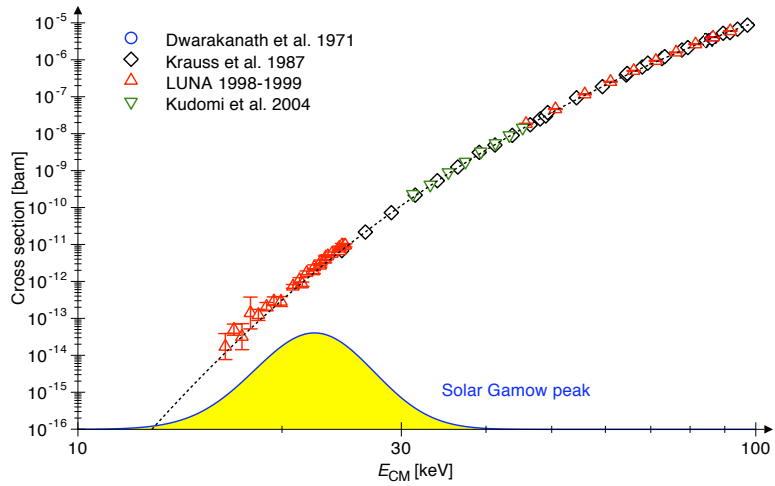}
}
\caption{\label{fig:figure6}Cross section of the $^{3}$He($^{3}$He,2p)$^{4}$He reaction. Data from LUNA \cite{Junker98-PRC,Bonetti99-PRL} and from other groups \cite{Dwarakanath71-PRC,Krauss87-NPA,Kudomi04-PRC}. The line is the extrapolation based on the measured $S(E)$-factor \cite{Bonetti99-PRL}.}
\end{figure}

The final set-up was made of eight 1 mm thick silicon detectors of 
5x5\,cm$^{2}$ area
placed around the beam inside the windowless  
target chamber filled with  $^{3}$He at the pressure of
0.5\,mbar.
The simultaneous detection of two protons has been 
the 
signature which unambiguously identified a
$^{3}$He($^{3}$He,2p)$^{4}$He
fusion reaction 
($Q$-value: 12.86 MeV).
Fig. \ref{fig:figure6} 
shows the results from LUNA \cite{Junker98-PRC,Bonetti99-PRL} together with 
higher energy measurements 
\cite{Dwarakanath71-PRC,Krauss87-NPA,Kudomi04-PRC} 
of the 
astrophysical factor $S(E)$.

For the first time a 
nuclear reaction has been measured in the laboratory at the
energy occurring in a star.
Its cross section varies by more than two orders of magnitude in the 
measured energy range. At the lowest energy of 16.5 keV it has 
the value of
0.02\,pb,
which corresponds to
a rate 
of about 2 events/month, rather low even for the "silent" 
experiments of underground physics.
No narrow resonance has been found within the solar Gamow peak and,
as a consequence, the astrophysical solution of the 
$^{8}$B and $^{7}$Be solar neutrino problem based on
its existence has been ruled out.

\subsection{$^{3}$He($^{4}$He,$\gamma$)$^7$Be: solar neutrino oscillations}

$^{3}$He($^4$He,$\gamma$)$^{7}$Be ($Q$-value: 1.586 MeV) is
the key reaction for the production of $^{7}$Be and $^{8}$B 
neutrinos in the Sun since their flux depends almost linearly on its
cross section.
Unless a recoil separator is used \cite{DiLeva09-PRL}, the cross section can be determined either from the detection of
the prompt $\gamma$ rays 
\cite{Holmgren59-PR,Parker63-PR,Nagatani69-NPA,Kraewinkel82-ZPA,Osborne82-PRL,
Alexander84-NPA,Hilgemeier88-ZPA}
or from the counting of the produced $^{7}$Be nuclei 
\cite{Osborne82-PRL,Robertson83-PRC,Volk83-ZPA,NaraSingh04-PRL,DiLeva09-PRL}.
The latter requires the detection of the 478 keV $\gamma$ due to the excited $^{7}$Li populated in the decay of $^{7}$Be
(half-life: 53.22 days). 

Both methods have been used in the past to determine the cross section 
in the energy range E$_{c.m.} \geq$ 107 keV 
but the S$_{3,4}$  extracted from the measurements of the 
induced $^{7}$Be activity was 13$\%$
higher than that obtained 
from the detection of the prompt $\gamma$-rays
\cite{Adelberger98-RMP}.

The underground experiment has been performed  
with the $^{4}$He$^{+}$ beam from the 400 kV accelerator in conjunction with a  windowless gas target
filled with $^{3}$He at 0.7 mbar.
The beam enters the target chamber 
and is stopped on the calorimeter (Fig. \ref{fig:figure3}).
The $^7$Be nuclei produced by the reaction
inside the $^3$He gas target are implanted into the calorimeter cap which,
after the irradiation, is removed and placed in front of a germanium detector
for the measurement of the $^7$Be activity. 

\begin{figure}
\centerline{
\includegraphics[width=0.75\linewidth]{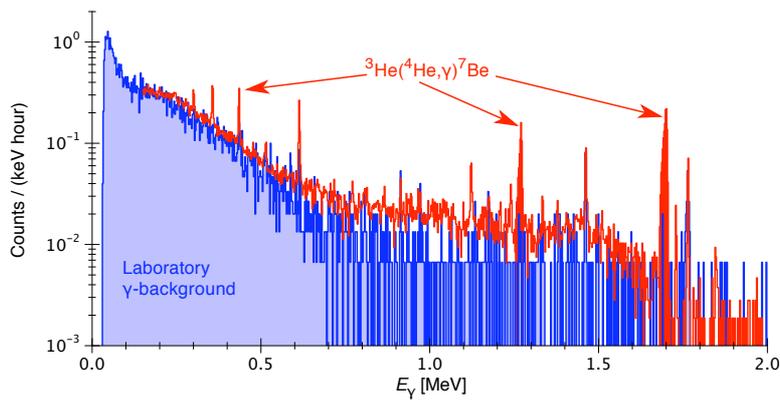}
}
\caption{\label{fig:figure7} $^3$He($^4$He,$\gamma$)$^7$Be spectrum at 250 keV beam energy (red) and the laboratory background (blue).}
\end{figure}

\begin{figure}
\centerline{
\includegraphics[width=0.75\linewidth]{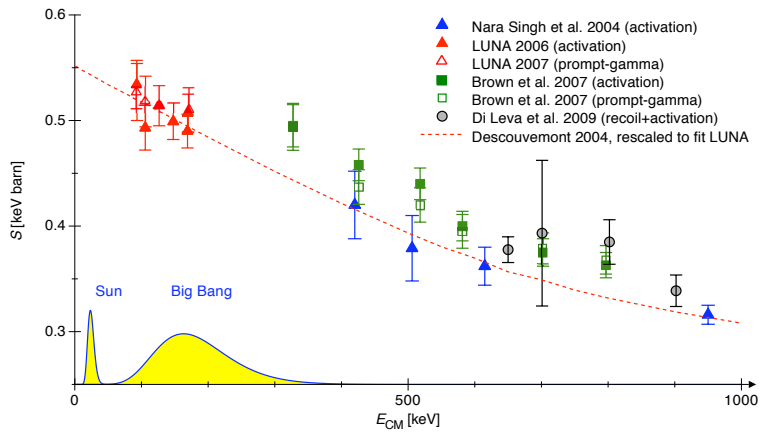}
}
\caption{\label{fig:figure8} Astrophysical S(E)-factor for $^3$He($^4$He,$\gamma$)$^7$Be. The results from the modern, high precision experiments are shown with their total error.}
\end{figure}

In the first phase of the experiment, the $^{3}$He($^4$He,$\gamma$)$^{7}$Be cross section has been obtained 
from the activation data \cite{Bemmerer06-PRL,Gyurky07-PRC} alone with a total uncertainty of about 4$\%$.
In the second phase, a new high accuracy measurement using simultaneously 
prompt and activation methods was performed down to the center of mass energy of 93 keV. 
The prompt capture $\gamma$-ray was detected by a 135$\%$
germanium heavily shielded and placed in close geometry with the target.
The spectrum taken at 250 keV beam energy is given in Fig. \ref{fig:figure7}.
The astrophysical factor obtained with the two methods \cite{Confortola07-PRC} is the same within the quoted experimental error (Fig.\ref{fig:figure8}).
Similar conclusions have then been reached in a new simultaneous activation and prompt experiment \cite{Brown07-PRC}
which covers the E$_{c.m.}$ energy range from 330 keV to 1230 keV.

The energy dependence of the cross section seems to be
theoretically well determined at low energy. If we leave the normalization as the only free parameter, we can rescale the fit of 
\cite{Descouvemont04-ADNDT} to our data and we obtain S$_{3,4}$(0)=0.560$\pm$0.017 keV barn.
Thanks to our small error, the total uncertainty on the $^{8}$B solar neutrino flux goes from 12 to 10$\%$, whereas the 
one on the $^{7}$Be flux goes from 9.4 to 5.5$\%$~\cite{Confortola07-PRC}. The $^{7}$Be flux is now theoretically predicted 
with an error as small as the experimental one which should soon be achieved by Borexino \cite{Borexino08-PRL}. Thanks to such small errors,
it will be possible to have a precise study of the signature typical of neutrino oscillations in matter, i.e. the energy dependence of the oscillation probability. 

The energy window covered by LUNA is above the solar Gamow peak but well within the Gamow peak of big-bang nucleosynthesis.
Our precise results clearly rule out the $^3$He($^4$He,$\gamma$)$^7$Be cross section as possible source of the discrepancy
between the predicted primordial $^7$Li abundance 
\cite{Coc04-ApJ}
and the much lower observed value \cite{Ryan00-ApJL,Bonifacio02-AA}.

\subsection{$^{14}$N(p,$\gamma$)$^{15}$O : the composition of the Sun and the age of the Universe}

$^{14}$N(p,$\gamma$)$^{15}$O ($Q$-value: 7.297 MeV)
is  the slowest reaction 
of the CNO cycle and it rules its energy production rate.
In particular, it is 
the key reaction to know
the $^{13}$N and $^{15}$O solar neutrino flux, which 
depends almost linearly on its cross section, as well as to determine 
the age of the globular clusters, which, consisting of 10$^{4}$-10$^{6}$ gravitationally bound
stars, are  
the oldest population of the galaxies. The luminosity of the turn-off point in the Hertzsprung-Russel diagram of a globular cluster, i.e. the point 
where the main sequence turns toward cooler and brighter stars, is used to determine the age of the cluster and to derive a lower limit on the age of the Universe
\cite{Krauss03-Science}.
A star
at the turn-off point is burning hydrogen in the shell through the CNO
cycle, this is the reason why the $^{14}N(p,\gamma)^{15}O$ cross section plays an important role 
in the age determination.

\begin{figure}
\centerline{
\includegraphics[width=0.75\linewidth]{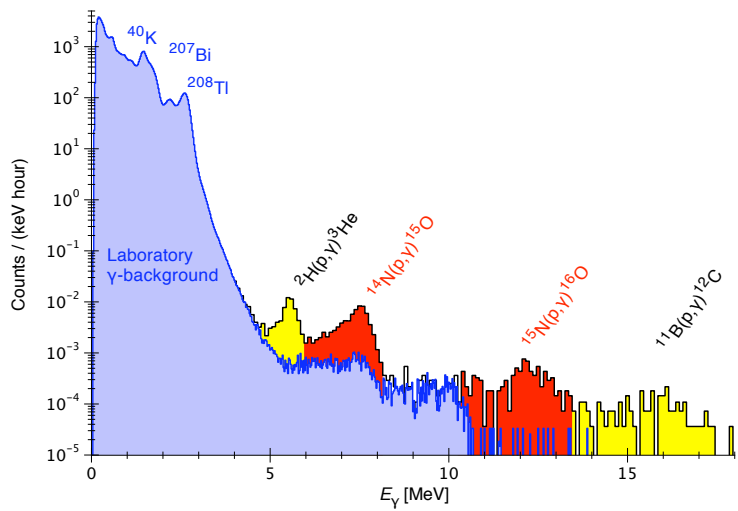}
}
\caption{\label{fig:figure9} BGO spectrum taken with 100 keV proton beam on Nitrogen of natural isotopic abundance. Red: peaks from $^{14}$N(p,$\gamma$)$^{15}$O and $^{15}$N(p,$\gamma$)$^{16}$O. 
Yellow: beam induced background. Blu: laboratory background. In spite of the small isotopic abundance of $^{15}$N (0.366$\%$ only) the peak due to $^{15}$N(p,$\gamma$)$^{16}$O
can be easily seen thanks to the much reduced background.}
\end{figure}

In the first phase of the LUNA study, data have been obtained 
down to 119 keV energy with
solid targets of TiN and a 126$ \% $ 
germanium detector.
This way, the five different radiative capture 
transitions which contribute to the $^{14}$N(p,$\gamma$)$^{15}$O
cross section at low energy were measured.
The total cross section was then studied
down to very low energy in the second phase of the experiment 
by using the 4$\pi$ BGO summing detector
placed around a
windowless gas target filled with nitrogen at 1 mbar pressure (the BGO spectrum at 100 keV beam energy is shown in Fig. \ref{fig:figure9}). At the lowest center of mass energy of 70 keV  a cross section of 0.24 pbarn was measured, with an event rate of
11 counts/day from the reaction.

\begin{figure}%
\centerline{
\includegraphics[width=0.75\linewidth]{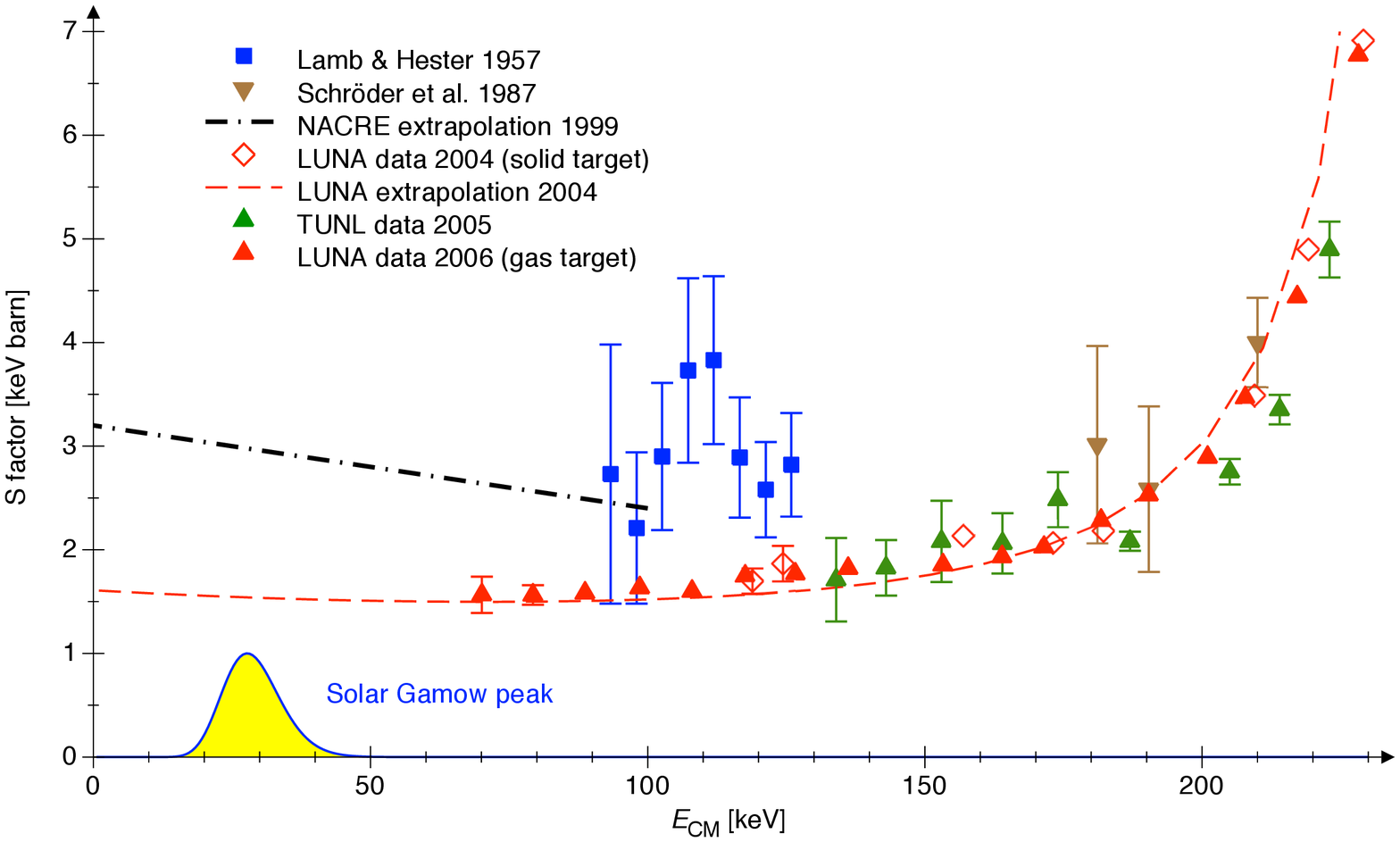}
}
\caption{\label{fig:figure10}
Astrophysical S(E)-factor of the $^{14}$N(p,$\gamma$)$^{15}$O reaction. The errors are statistical only (the systematic ones are similar).}
\end{figure}

The results obtained first with  the germanium detector ~\cite{Formicola04-PLB,Imbriani05-EPJA} 
and then with
the BGO set-up~\cite{Lemut06-PLB} were about a factor two lower than the existing extrapolation 
\cite{CF88-ADNDT,Adelberger98-RMP,NACRE99-NPA}
from previous data
\cite{Lamb57-PR,Schroeder87-NPA}
at very low energy (Fig.\ref{fig:figure10}), while in agreement with results from indirect methods \cite{Bertone01-PRL,Bertone02-PRC,Mukhamedzhanov03-PRC,Nelson03-PRC,Yamada04-PLB,Schuermann08-PRC}.

As a consequence, the CNO neutrino yield
in the Sun is decreased by about a factor two, and the age of the
globular clusters is increased by 0.7-1 billion years \cite{Imbriani04-AA}
up to 14 billion years. 
The lower cross section is affecting also stars which are much more 
evolved than our Sun: in particular, 
the dredge-up of carbon to the surface of asymptotic giant branch stars \cite{Herwig05-ARAA} is more efficient \cite{Herwig04-APJ}. 

The main conclusion from the LUNA data has been confirmed by an independent study at higher energy 
~\cite{Runkle05-PRL}. However, there is a 15$\%$ difference 
between the total S-factor extrapolated by the two experiments at the Gamow peak of the Sun. 
In particular, this difference arises from the extrapolation of the capture to the ground state in $^{15}$O,
a transition strongly affected by interference effects between several resonances and the direct capture
mechanism. 

In order to provide precise data for the ground state capture, a third phase of
the  $^{14}$N(p,$\gamma$)$^{15}$O study has been performed with a composite germanium detector in the beam energy region immediately above the 259 keV resonance, where precise data 
effectively constrain a fit for the ground state transition in the R-matrix \cite{Lane58-RMP} framework. 
This way the total error on the S-factor was
reduced to 8$\%$:
S$_{1,14}$(0)=1.57$\pm$0.13 keV barn~\cite{Marta08-PRC}. This is significant because,
finally solved the solar neutrino problem, we are now facing the solar composition problem:
the conflict between helioseismology and the new metal abundances that emerged from improved modeling 
of the photosphere~\cite{PenaGaray08-arxiv,Haxton08-ApJ}.  Thanks to the relatively small error, it will soon be possible to
measure the metallicity of the core of the Sun (i.e. the contents of elements different from hydrogen and helium)
by comparing the detected CNO neutrino flux with the predicted one. 
As a matter of fact, the CNO neutrino flux is decreased by about 35$\%$ in going from
the high to the low metallicity scenario. This way it will be possible to test 
whether the early Sun was chemically homogeneous ~\cite{Haxton08-arxiv}, a
key assumption of the standard Solar Model.

\subsection{Ongoing measurements}

The solar phase of LUNA has almost reached the end. A new and rich program of nuclear astrophysics
mainly devoted to
the Mg-Al and Ne-Na cycles has already started at the 400 kV facility about 2 years ago 
with the measurement of 
$^{25}$Mg(p,$\gamma$)$^{26}$Al \cite{Costantini09-RPP}.
These cycles become important for second generation stars with central temperatures and masses higher than those of our Sun \cite{Rolfs88-Book, Iliadis07-Book}. Due to the higher Coulomb 
barriers, these cycles are relatively unimportant for energy generation while being essential for the nucleosynthesis of elements with mass number higher than 20.
Low energy resonances (or the low energy part of the direct capture) inaccessible in a laboratory at the Earth surface, could become measurable underground. Some of the selected reactions have already been measured over ground but an underground re-investigation can substantially improve the knowledge of the related reaction rate in the different astrophysical scenarios responsible, in particular, of the texture of the isotopes which are filling the Universe.

LUNA is now measuring 
$^{15}$N(p,$\gamma$)$^{16}$O with enriched $^{15}$N targets. $^{15}$N(p,$\gamma$)$^{16}$O is the leak reaction from the first to the second CNO cycle. 
The results already obtained with nitrogen of natural isotopic composition (0.366\% $^{15}$N) \cite{Bemmerer09-JPG} extend to energies lower than ever measured
before and 
provide a cross section which is about a factor two lower than previously believed at novae energies. 

The measurement under preparation is not connected to the hydrogen burning cycles but it is a key reaction of big-bang nucleosynthesis: $^{2}$H($^{4}$He, $\gamma$)$^{6}$ Li. As a matter of fact, such reaction determines the amount of primordial $^{6}$Li in the Universe. Recently, the $^{6}$Li isotope has been detected in a number of metal poor stars \cite{Smith93-ApJ, Asplund06-ApJ} and its quantity has been found to be 2-3 orders of magnitude higher than what expected from BBN \cite{Coc04-ApJ}.

% =======================
\section{Outlook}

As we have seen, the approach to measure cross sections directly at the energy of astrophysical interest has been used with great success for the reactions governing stellar hydrogen-burning. In order to keep pace with the rapid progress of observational astronomy and astrophysical modeling, the same should now be done for the nuclear reactions governing helium and carbon burning and
producing inside stars the neutrons which give rise to the so-called astrophysical s-process. Owing to the different nature of these reactions, new techniques and new equipment are necessary for this task.

The $^{12}$C($\alpha$,$\gamma$)$^{16}$ O reaction is often referred to as the ""Holy Grail"" of nuclear astrophysics \cite{Weaver93-PhysRep,Wallerstein97-RMP,Buchmann06-NPA,Woosley07-PhysRep,Tur07-ApJ,Tang07-PRL}. It plays a fundamental role in the evolution of stars during the helium-burning phase and determines the abundance ratio between carbon and oxygen, the two elements of fundamental importance to the development of life. This abundance ratio, in turn, influences the nucleosynthesis of elements up to the iron peak for massive stars, the cooling timescale of white dwarfs and the properties of thermonuclear as well as core collapse supernovae.

\begin{table}[h]
\caption{Nuclear reactions of astrophysical interest recommended for study at future underground accelerator facilities.}
\label{table:reactions}
\resizebox{\textwidth}{!}{
\begin{tabular}{|l|l|r|r|l|}
\hline
Type & Reaction & $Q$-value [MeV] & $E_{\rm Gamow}$ [keV] & Reference \\[2mm] \hline
($\alpha$,$\gamma$) & $^{2}$H($\alpha$,$\gamma$)$^{6}$Li & 1.5 & 100-500 & \cite{Caciolli09-EPJA} \\
 & $^{12}$C($\alpha$,$\gamma$)$^{16}$O & 7.2 & 300 & \cite{Schuermann05-EPJA,Matei06-PRL,Tang07-PRL} \\
 & $^{14}$N($\alpha$,$\gamma$)$^{18}$F & 4.4 & 200-700 & \cite{Goerres00-PRC} \\ %
 & $^{15}$N($\alpha$,$\gamma$)$^{19}$F &  4.0 & 500 & \cite{Wilmes02-PRC} \\
$^{12}$C-induced & $^{12}$C($^{12}$C,$\alpha$)$^{20}$Ne & 4.6 & 1500 & \cite{Spillane07-PRL,Caciolli09-EPJA} \\
 & $^{12}$C($^{12}$C,p)$^{23}$Na & 2.2 & 1500 & \cite{Spillane07-PRL,Caciolli09-EPJA} \\
($\alpha$,n) & $^{13}$C($\alpha$,n)$^{16}$O & 2.2 & 200 & \cite{Harissopulos05-PRC,Heil08-PRC} \\
  & $^{22}$Ne($\alpha$,n)$^{25}$Mg & -0.5 & 500  &  \cite{Jaeger01-PRL}\\
\hline
\end{tabular}
}%resizebox
\end{table}

The $^{12}$C+$^{12}$C fusion reactions form the onset of carbon burning. Their rate determines the evolution of massive stars up to a modest end as a white dwarf or a fiery death as core-collapse supernovae \cite{Woosley02-RMP, Hillebrandt00-ARAA}. It also affects the ignition conditions and time scales of thermonuclear supernovae. Recent studies, while uncovering interesting facts \cite{Barron06-NPA,Spillane07-PRL}, stopped short of the astrophysical energy range due to the high laboratory background.

The $^{13}$C($\alpha$,n)$^{16}$O and $^{22}$Ne($\alpha$,n)$^{25}$Mg  reactions provide the neutrons for the build-up of the s-process isotopes \cite{Cameron57-PASP,Burbidge57-RMP, Wallerstein97-RMP,Straniero06-NPA,Heil08-PRC}. Most of the elements heavier than iron are produced through this process, which involves a series of subsequent neutron captures and $\beta$-decays.

In order to address these exciting cases, it has been called for the installation of a new accelerator with a larger voltage up to a few MV in a deep underground laboratory \cite{Nupecc05-Roadmap}. In addition to protons and $\alpha$-particles, it should also be able to accelerate ions up to carbon or even oxygen, so that also carbon burning can be studied. As to the beam current, for all these ions an intensity of typically 1 mA will be required for the necessary sensitivity.

Such a project is presently under discussion at several different sites in Europe and North America. It would be natural to install a new underground accelerator at the site of the present LUNA machine, the LNGS laboratory, with its excellent infrastructure and proven low background. However, due to limited underground laboratory space there, also possible other sites must be actively studied. 

One recent example is the Canfranc laboratory in Spain \cite{Barcelona09-www}, which is less well-shielded (2400 m.w.e.) than Gran Sasso and therefore has a higher remaining muon flux. However, at the present LUNA site the background is dominated by neutrons, not muons, \cite{Bemmerer05-EPJA}, and the neutron flux at the two sites is actually comparable. The access to this site is by horizontal road tunnel, facilitating the installation of complex, maintenance-intensive equipment like an accelerator. 

In the United States, two accelerators with connected beamlines are included in the DUSEL underground science facility planned at the site of the previous Homestake experiment \cite{DOE08-arxiv}. One machine should have similar tasks as the present LUNA 400\,kV accelerator, but with greatly increased beam intensity, whereas the second accelerator should be in the MV range and address the science cases described above. 

A different approach is followed in two projects in Boulby (United Kingdom, 2800 m.w.e.) and Praid (Romania, 900 m.w.e.) \cite{Strieder08-NPA3,Bordeanu08-NPA3}: Here the laboratory should be placed in a salt matrix deep underground, with its generally much lower levels of uranium and thorium and, therefore, also lower neutron flux. 

Regardless of the outcome of the ongoing siting discussion, both astrophysics and nuclear physics will greatly benefit from the new precision that will be enabled by a future, higher-voltage accelerator underground. A better understanding of nuclear burning in stars by direct new data will allow to model stellar scenarios that are now understood only in general terms. In addition, improved experimental data for nuclear fusion reactions near or below the Coulomb barrier will open a fresh challenge for theoretical modeling of these reactions. 

% =======================
\section{Conclusions}
LUNA started underground nuclear astrophysics twenty years ago  in the core of Gran Sasso, below 1400 meters of
dolomite rock. The extremely low background has allowed for nuclear physics experiments with very small count rate, down to a few events per year. The important reactions responsible for the hydrogen burning in the Sun have been studied for the first time down to the relevant stellar energies. 
As a consequence, fifty years after the first pioneering cross section measurements, nuclear physics is not anymore an important error source of the solar model and solar neutrinos can now be exploited to probe the deep interior of the Sun. When applied to astrophysical scenarios different from the Sun, LUNA results increase the limit on the age of the Universe up to 14 billion  years. 

LUNA has already experienced the important progress achievable in the comprehension of the hydrogen burning thanks to the underground environment. In the next two decades underground nuclear astrophysics will try to reach similar results in the study of the helium and carbon burning and of the neutron sources in the stars. 

\section*{Acknowledgments}

It is a pleasure to thank our colleagues of LUNA without whom the results reviewed in this paper would have not been obtained.

% ==================
% ==================

% =======================


\begin{thebibliography}{100}

\bibitem{Eddington20-Nature}
Eddington A,
\newblock Nature 16:14 (1920).

\bibitem{Clayton84-Book}
Clayton DD,
\newblock {\em Principles of Stellar Evolution and Nucleosynthesis} (University
  of Chicago Press, 1984).

\bibitem{Rolfs88-Book}
Rolfs C, Rodney W,
\newblock {\em Cauldrons in the Cosmos} (University of Chicago Press, Chicago,
  1988).

\bibitem{Clayton03-Book}
Clayton DD,
\newblock {\em Handbook of Isotopes in the Cosmos: Hydrogen to Gallium}
  (Cambridge University Press, Cambridge, 2003).

\bibitem{Iliadis07-Book}
Iliadis C,
\newblock {\em Nuclear Physics of Stars} (Wiley-VCH, Weinheim, 2007).

\bibitem{Salpeter54-AustralJP}
{Salpeter} EE,
\newblock Australian Journal of Physics 7:373 (1954).

\bibitem{Assenbaum87-ZPA}
Assenbaum H, Langanke K, Rolfs C,
\newblock Z.~Phys.~A 327:461 (1987).

\bibitem{Raiola04-EPJA}
{Raiola {\it et al.}} F,
\newblock Eur.~Phys.~J.~A 19:283 (2004).

\bibitem{Huke08-PRC}
{Huke} A, et~al.,
\newblock Phys.~Rev.~C 78:015803 (2008).

\bibitem{Gilmore08-Book}
Gilmore G,
\newblock {\em {{Practical $\gamma$-ray spectrometry, 2nd edition}}} (John
  Wiley and Sons, New York, 2008).

\bibitem{Koehler09-Apradiso}
K\"ohler M, et~al.,
\newblock Appl.~Radiat.~Isot. 67:736 (2009).

\bibitem{Caciolli09-EPJA}
Caciolli A, et~al.,
\newblock Eur.~Phys.~J.~A 39:179 (2009).

\bibitem{Laubenstein04-Apradiso}
Laubenstein M, et~al.,
\newblock Appl.~Radiat.~Isot. 61:167 (2004).

\bibitem{MACRO90-PLB}
{Ahlen} SP, et~al.,
\newblock Phys.~Lett.~B 249:149 (1990).

\bibitem{Bemmerer05-EPJA}
Bemmerer D, et~al.,
\newblock Eur.~Phys.~J.~A 24:313 (2005).

\bibitem{Szucs10-EPJA}
{Sz{\"u}cs} T, et~al.,
\newblock Eur.~Phys.~J.~A 44:513 (2010).

\bibitem{Wulandari04-APP}
{Wulandari} H, {Jochum} J, {Rau} W, {von Feilitzsch} F,
\newblock Astropart. Phys. 22:313 (2004).

\bibitem{Formaggio04-ARNPS}
Formaggio JA, Martoff C,
\newblock Annu. Rev. Nucl. Part. Sci. 54:361 (2004).

\bibitem{Bethe39-PR_letter}
Bethe H,
\newblock Phys.~Rev. 55:103 (1939).

\bibitem{Weizsaecker38-PZ}
von Weizs\"acker CF,
\newblock Phys.~Z. 39:633 (1938).

\bibitem{KamLAND05-Nature}
{Araki} T, et~al.,
\newblock Nature 436:499 (2005).

\bibitem{Gaisser02-ARNPS}
{Gaisser} TK, {Honda} M,
\newblock Annu. Rev. Nucl. Part. Sci. 52:153 (2002), arXiv:hep-ph/0203272.

\bibitem{Elliott02-ARNPS}
{Elliott} SR, {Vogel} P,
\newblock Annu. Rev. Nucl. Part. Sci. 52:115 (2002), arXiv:hep-ph/0202264.

\bibitem{Perkins84-ARNPS}
Perkins DH,
\newblock Annu. Rev. Nucl. Part. Sci. 34:1 (1984).

\bibitem{Gaitskell04-ARNPS}
{Gaitskell} RJ,
\newblock Annu. Rev. Nucl. Part. Sci. 54:315 (2004).

\bibitem{Bettini07-arxiv}
Bettini A,
\newblock J.~Phys.~Conf.~Ser. 120:082001 (2008), arXiV:0712.1051.

\bibitem{Junker98-PRC}
Junker M, et~al.,
\newblock Phys.~Rev.~C 57:2700 (1998).

\bibitem{Bonetti99-PRL}
Bonetti R, et~al.,
\newblock Phys.~Rev.~Lett. 82:5205 (1999).

\bibitem{Casella02-NPA}
Casella C, et~al.,
\newblock Nucl.~Phys.~A 706:203 (2002).

\bibitem{Formicola04-PLB}
Formicola A, et~al.,
\newblock Phys.~Lett.~B 591:61 (2004).

\bibitem{Imbriani05-EPJA}
Imbriani G, et~al.,
\newblock Eur.~Phys.~J.~A 25:455 (2005).

\bibitem{Lemut06-PLB}
Lemut A, Bemmerer D, et~al.,
\newblock Phys.~Lett.~B 634:483 (2006).

\bibitem{Bemmerer06-NPA}
Bemmerer D, et~al.,
\newblock Nucl.~Phys.~A 779:297 (2006).

\bibitem{Marta08-PRC}
{Marta} M, et~al.,
\newblock Phys.~Rev.~C 78:022802(R) (2008).

\bibitem{Confortola07-PRC}
{Confortola} F, et~al.,
\newblock Phys.~Rev.~C 75:065803 (2007).

\bibitem{Costantini08-NPA}
Costantini H, et~al.,
\newblock Nucl.~Phys.~A 814:144 (2008).

\bibitem{Bemmerer09-JPG}
Bemmerer D, et~al.,
\newblock J.~Phys.~G 36:045202 (2009).

\bibitem{Bemmerer06-PRL}
{Bemmerer} D, et~al.,
\newblock Phys.~Rev.~Lett. 97:122502 (2006).

\bibitem{Gyurky07-PRC}
{Gy{\"u}rky} G, et~al.,
\newblock Phys.~Rev.~C 75:035805 (2007).

\bibitem{Greife94-NIMA}
Greife U, et~al.,
\newblock Nucl.~Inst.~Meth.~A 350:327 (1994).

\bibitem{Formicola03-NIMA}
Formicola A, et~al.,
\newblock Nucl.~Inst.~Meth.~A 507:609 (2003).

\bibitem{Costantini00-PLB}
{Costantini} H, et~al.,
\newblock Phys.~Lett.~B 482:43 (2000).

\bibitem{Casella02-NIMA}
Casella C, et~al.,
\newblock Nucl.~Inst.~Meth.~A 489:160 (2002).

\bibitem{Marta06-NIMA}
Marta M, et~al.,
\newblock Nucl.~Inst.~Meth.~A 569:727 (2006).

\bibitem{Goerres80-NIM}
G{\"o}rres J, et~al.,
\newblock Nucl.~Inst.~Meth. 177:295 (1980).

\bibitem{Allison58-RMP}
Allison SK,
\newblock Rev.~Mod.~Phys. 30:1138 (1958).

\bibitem{Elekes03-NIMA}
Elekes Z, et~al.,
\newblock Nucl.~Inst.~Meth.~A 503:580 (2003).

\bibitem{Bahcall89-Book}
{Bahcall} JN,
\newblock {\em {Neutrino astrophysics}} (Cambridge and New York, Cambridge
  University Press, 1989).

\bibitem{Castellani97-PhysRep}
{Castellani} V, et~al.,
\newblock Phys.~Rep. 281:309 (1997), arXiv:astro-ph/9606180.

\bibitem{PenaGaray08-arxiv}
{Pe\~na-Garay} C, {Serenelli} A,
\newblock ArXiv e-prints  (2008), 0811.2424.

\bibitem{Bahcall64-PRL}
{Bahcall} JN,
\newblock Phys.~Rev.~Lett. 12:300 (1964).

\bibitem{Davis64-PRL}
{Davis} R,
\newblock Phys.~Rev.~Lett. 12:303 (1964).

\bibitem{Weinberg72-Book}
Weinberg S,
\newblock {\em Gravitation and Cosmology} (John Wiley and Sons, 1972).

\bibitem{Griffiths63-CJP}
Griffiths GM, Lal M, Scarfe CD,
\newblock Can.~J.~Phys. 41:724 (1963).

\bibitem{Schmid96-PRL}
{Schmid} GJ, et~al.,
\newblock Phys.~Rev.~Lett. 76:3088 (1996).

\bibitem{Marcucci06-NPA}
Marcucci L, Nollett K, Schiavilla R, Wiringa R,
\newblock Nucl.~Phys.~A 777:111 (2006).

\bibitem{Fowler72-NATURE}
{Fowler} WA,
\newblock Nature 238:24 (1972).

\bibitem{Fetisov72-PLB}
{Fetisov} VN, {Kopysov} YS,
\newblock Phys.~Lett.~B 40:602 (1972).

\bibitem{Dwarakanath71-PRC}
{Dwarakanath} MR, {Winkler} H,
\newblock Phys.~Rev.~C 4:1532 (1971).

\bibitem{Krauss87-NPA}
{Krauss} A, {Becker} HW, {Trautvetter} HP, {Rolfs} C,
\newblock Nucl.~Phys.~A 467:273 (1987).

\bibitem{Kudomi04-PRC}
{Kudomi} N, et~al.,
\newblock Phys.~Rev.~C 69:015802 (2004), arXiv:astro-ph/0306454.

\bibitem{DiLeva09-PRL}
{di Leva} A, et~al.,
\newblock Phys.~Rev.~Lett. 102:232502 (2009).

\bibitem{Holmgren59-PR}
{Holmgren} HD, {Johnston} RL,
\newblock Phys.~Rev. 113:1556 (1959).

\bibitem{Parker63-PR}
Parker P, Kavanagh R,
\newblock Phys.~Rev. 131:2578 (1963).

\bibitem{Nagatani69-NPA}
Nagatani K, Dwarakanath M, Ashery D,
\newblock Nucl.~Phys.~A 128:325 (1969).

\bibitem{Kraewinkel82-ZPA}
Kr\"awinkel H, et~al.,
\newblock Z.~Phys.~A 304:307 (1982).

\bibitem{Osborne82-PRL}
Osborne J, et~al.,
\newblock Phys.~Rev.~Lett. 48:1664 (1982).

\bibitem{Alexander84-NPA}
Alexander T, Ball G, Lennard W, Geissel H,
\newblock Nucl.~Phys.~A 427:526 (1984).

\bibitem{Hilgemeier88-ZPA}
Hilgemeier M, et~al.,
\newblock Z.~Phys.~A 329:243 (1988).

\bibitem{Robertson83-PRC}
Robertson R, et~al.,
\newblock Phys.~Rev.~C 27:11 (1983).

\bibitem{Volk83-ZPA}
Volk H, Kr\"awinkel H, Santo R, Wallek L,
\newblock Z.~Phys.~A 310:91 (1983).

\bibitem{NaraSingh04-PRL}
{Nara Singh} B, Hass M, Nir-El Y, Haquin G,
\newblock Phys.~Rev.~Lett. 93:262503 (2004).

\bibitem{Adelberger98-RMP}
Adelberger E, et~al.,
\newblock Rev.~Mod.~Phys. 70:1265 (1998).

\bibitem{Brown07-PRC}
{Brown} TAD, et~al.,
\newblock Phys.~Rev.~C 76:055801 (2007), 0710.1279.

\bibitem{Descouvemont04-ADNDT}
Descouvemont P, et~al.,
\newblock At. Data Nucl. Data Tables 88:203 (2004).

\bibitem{Borexino08-PRL}
Arpesella C, et~al.,
\newblock Phys.~Rev.~Lett. 101:091302 (2008).

\bibitem{Coc04-ApJ}
Coc A, et~al.,
\newblock Astrophys.~J. 600:544 (2004).

\bibitem{Ryan00-ApJL}
Ryan S, et~al.,
\newblock Astrophys.~J. 530:L57 (2000).

\bibitem{Bonifacio02-AA}
Bonifacio P, et~al.,
\newblock Astron.~Astrophys. 390:91 (2002).

\bibitem{Krauss03-Science}
Krauss L, Chaboyer B,
\newblock Science 299:65 (2003).

\bibitem{CF88-ADNDT}
Caughlan G, Fowler W,
\newblock At. Data Nucl. Data Tables 40:283 (1988).

\bibitem{NACRE99-NPA}
Angulo C, et~al.,
\newblock Nucl.~Phys.~A 656:3 (1999).

\bibitem{Lamb57-PR}
Lamb W, Hester R,
\newblock Phys.~Rev. 108:1304 (1957).

\bibitem{Schroeder87-NPA}
Schr{\"o}der U, et~al.,
\newblock Nucl.~Phys.~A 467:240 (1987).

\bibitem{Bertone01-PRL}
Bertone P, et~al.,
\newblock Phys.~Rev.~Lett. 87:152501 (2001).

\bibitem{Bertone02-PRC}
Bertone PF, et~al.,
\newblock Phys.~Rev.~C 66:055804 (2002).

\bibitem{Mukhamedzhanov03-PRC}
Mukhamedzhanov A, et~al.,
\newblock Phys.~Rev.~C 67:065804 (2003).

\bibitem{Nelson03-PRC}
Nelson SO, et~al.,
\newblock Phys.~Rev.~C 68:065804 (2003).

\bibitem{Yamada04-PLB}
Yamada K, et~al.,
\newblock Phys.~Lett.~B 579:265 (2004).

\bibitem{Schuermann08-PRC}
Sch\"urmann D, et~al.,
\newblock Phys.~Rev.~C 77:055803 (2008).

\bibitem{Imbriani04-AA}
Imbriani G, et~al.,
\newblock Astron.~Astrophys. 420:625 (2004).

\bibitem{Herwig05-ARAA}
{Herwig} F,
\newblock Ann. Rev. Astron. Astroph. 43:435 (2005).

\bibitem{Herwig04-APJ}
{Herwig} F, {Austin} SM,
\newblock Astrophys.~J. 613:L73 (2004).

\bibitem{Runkle05-PRL}
Runkle RC, et~al.,
\newblock Phys.~Rev.~Lett. 94:082503 (2005).

\bibitem{Lane58-RMP}
Lane AM, Thomas RG,
\newblock Rev.~Mod.~Phys. 30:257 (1958).

\bibitem{Haxton08-ApJ}
{Haxton} WC, {Serenelli} AM,
\newblock Astrophys.~J. 687:678 (2008).

\bibitem{Haxton08-arxiv}
{Haxton} W,
\newblock Journal of Physics Conference Series 173:012014 (2009), 0809.3342.

\bibitem{Costantini09-RPP}
{Costantini} H, et~al.,
\newblock Rep.~Prog.~Phys. 72:086301 (2009).

\bibitem{Smith93-ApJ}
Smith V, Lambert D, Nissen P,
\newblock Astrophys.~J. 408:262 (1993).

\bibitem{Asplund06-ApJ}
{Asplund} M, et~al.,
\newblock Astrophys.~J. 644:229 (2006).

\bibitem{Weaver93-PhysRep}
{Weaver} TA, {Woosley} SE,
\newblock Phys.~Rep. 227:65 (1993).

\bibitem{Wallerstein97-RMP}
{Wallerstein} G, et~al.,
\newblock Rev.~Mod.~Phys. 69:995 (1997).

\bibitem{Buchmann06-NPA}
Buchmann L, Barnes C,
\newblock Nucl.~Phys.~A 777:254 (2006).

\bibitem{Woosley07-PhysRep}
{Woosley} SE, {Heger} A,
\newblock Phys.~Rep. 442:269 (2007), arXiv:astro-ph/0702176.

\bibitem{Tur07-ApJ}
{Tur} C, {Heger} A, {Austin} SM,
\newblock Astrophys.~J. 671:821 (2007), 0705.4404.

\bibitem{Tang07-PRL}
{Tang} XD, et~al.,
\newblock Phys.~Rev.~Lett. 99:052502 (2007).

\bibitem{Schuermann05-EPJA}
{Sch{\"u}rmann} D, et~al.,
\newblock European Physical Journal A 26:301 (2005).

\bibitem{Matei06-PRL}
{Matei} C, et~al.,
\newblock Phys.~Rev.~Lett. 97:242503 (2006).

\bibitem{Goerres00-PRC}
{G\"orres} J, et~al.,
\newblock Phys.~Rev.~C 62:055801 (2000).

\bibitem{Wilmes02-PRC}
Wilmes S, et~al.,
\newblock Phys.~Rev.~C 66:065802 (2002).

\bibitem{Spillane07-PRL}
Spillane T, et~al.,
\newblock Phys.~Rev.~Lett. 98:122501 (2007).

\bibitem{Harissopulos05-PRC}
{Harissopulos} S, et~al.,
\newblock Phys.~Rev.~C 72:062801 (2005), arXiv:nucl-ex/0509014.

\bibitem{Heil08-PRC}
{Heil} M, et~al.,
\newblock Phys.~Rev.~C 78:025803 (2008).

\bibitem{Jaeger01-PRL}
{Jaeger} M, {Kunz} R, {Mayer} A, {Hammer} JW, {Staudt} G, et~al.,
\newblock Phys.~Rev.~Lett. 87:202501 (2001).

\bibitem{Woosley02-RMP}
{Woosley} SE, {Heger} A, {Weaver} TA,
\newblock Rev.~Mod.~Phys. 74:1015 (2002).

\bibitem{Hillebrandt00-ARAA}
{Hillebrandt} W, {Niemeyer} JC,
\newblock Ann. Rev. Astron. Astroph. 38:191 (2000), arXiv:astro-ph/0006305.

\bibitem{Barron06-NPA}
Barr\'on-Palos L, et~al.,
\newblock Nucl.~Phys.~A 779:318 (2006).

\bibitem{Cameron57-PASP}
{Cameron} AGW,
\newblock Publ. Astron. Soc. Pacific 69:201 (1957).

\bibitem{Burbidge57-RMP}
{Burbidge} EM, {Burbidge} GR, {Fowler} WA, {Hoyle} F,
\newblock Rev.~Mod.~Phys. 29:547 (1957).

\bibitem{Straniero06-NPA}
{Straniero} O, {Gallino} R, {Cristallo} S,
\newblock Nucl.~Phys.~A 777:311 (2006), arXiv:astro-ph/0501405.

\bibitem{Nupecc05-Roadmap}
{Nuclear Physics European Collaboration Committee (NuPECC), Roadmap 2005},
\newblock {available at {\tt http://www.nupecc.org/pub/NuPECC\_Roadmap.pdf}}.

\bibitem{Barcelona09-www}
{{Workshop on Nuclear Astrophysics Opportunities at the Underground Laboratory
  in Canfranc}},
\newblock {Barcelona/Spain}, 2009.

\bibitem{DOE08-arxiv}
{{DOE/NSF Nuclear Science Advisory Committee}},
\newblock (2008), arXiV:0809.3137.

\bibitem{Strieder08-NPA3}
Strieder F,
\newblock J.~Phys.~G 35:014009 (2008).

\bibitem{Bordeanu08-NPA3}
Bordeanu C, et~al.,
\newblock J.~Phys.~G 35:014011 (2008).

\end{thebibliography}
\end{document}